\documentclass[letterpaper]{article} 
\usepackage{aaai2026}  
\nocopyright
\usepackage{times}  
\usepackage{helvet}  
\usepackage{courier}  
\usepackage[hyphens]{url}  
\usepackage{graphicx} 
\usepackage{natbib}  
\usepackage{caption} 
\usepackage{algorithm}
\usepackage{algorithmic}

\usepackage{newfloat}
\usepackage{listings}
\DeclareCaptionStyle{ruled}{labelfont=normalfont,labelsep=colon,strut=off} 
\floatstyle{ruled}
\newfloat{listing}{tb}{lst}{}
\floatname{listing}{Listing}
\title{From Theory of Mind to Theory of Environment: Counterfactual Simulation of Latent Environmental Dynamics}
\author {
    Ryutaro Uchiyama
}
\affiliations{
    Singapore University of Technology and Design\\
    \texttt{ryutaro\_uchiyama@sutd.edu.sg}
}

\usepackage{bibentry}

\begin{document}

\maketitle

\begin{abstract}
The vertebrate motor system employs dimensionality-reducing strategies to limit the complexity of movement coordination, for efficient motor control. But when environments are dense with hidden action–outcome contingencies, movement complexity can promote behavioral innovation. Humans, perhaps uniquely, may infer the presence of hidden environmental dynamics from social cues, by drawing upon computational mechanisms shared with Theory of Mind. This proposed ``Theory of Environment'' supports behavioral innovation by expanding the dimensionality of motor exploration.
\end{abstract}


\section{The problem of behavioral innovation}

The flexibility and creativity of human behavior remain an enigma. Theories of cultural evolution explain how the emergence of conformism and imitation enabled behavioral innovations to persist and cumulate across generations, resulting in the ecological success of our species \cite{Boyd_Richerson_1985}. Behavioral innovation itself, however, remains poorly understood, often relying on assumptions of random variation that are formally analogous to genetic mutation. Here we propose a novel socio-cognitive mechanism, grounded in Theory of Mind computation \cite{Barnby_Alon_Bellucci_Schilbach_Frith_Bell_2024}, that helps bridge this explanatory gap.

The human brain controls approximately 600 muscles and 350 joints to generate desirable outcomes in a 3-dimensional space, yielding a highly redundant system in which a given action objective can be realized by a vast number of possible motor configurations. To reduce this sprawling complexity \cite{Bernstein_1967}, the vertebrate motor system organizes muscular activation into coordinated “muscle synergies” \cite{Overduin_dAvella_Roh_Bizzi_2008} that impose strategic low-dimensional constraints onto high-dimensional biomechanics. By constraining the variability of movement coordination, muscle synergies facilitate efficient whole-body control, but necessarily limit the exploration of movement-coordination structures and thus possible behaviors. Open-ended behavioral exploration is generally a costly investment, as evolution optimizes for multiplicative (i.e., geometric mean) fitness, where a single zero-fitness episode wipes out all prior gains. Assumptions of additive utility in reinforcement learning hence underestimate this vulnerability to exploration risk. The restricted behavioral repertoire of non-human primates \cite{Tennie_Call_Tomasello_2009} should be construed not as a functional deficit, but as reflecting a general solution to the problem of motor complexity. 

Recently, leading research groups in human evolutionary biology \cite{Morgan_Feldman_2024} and computational cognitive science \cite{Chu_Tenenbaum_Schulz_2024} have independently argued that the species-unique feature of human behavior is its open-ended variability. Such claims suggest that humans may have innovated the means to ``unbind'' acquired constraints on movement degrees-of-freedom -- thus becoming able to not only reduce but also \textit{expand} motor exploration complexity. Recent approaches in the movement sciences illustrate how such increases in the dimensional complexity of motor coordination can facilitate skill acquisition \cite{Dhawale_Smith_Olveczky_2017}. 

Real ecological environments typically contain an unbounded number of hidden action–outcome contingencies (i.e., environmental dynamics) that can be potentially unlocked by skill acquisition -- constituting an open-ended search space. The density of these latent environmental goals (``teleological depth'') thus determines the scope of prospective future gains in the \textit{controllability} of environmental outcomes \cite{Ligneul_Mainen_Ly_Cools_2022, Mancinelli_Roiser_Dayan_2021}. Such untapped prospective goal-states can offset the investment cost of behavioral exploration \cite{Molinaro_Colas_Oudeyer_Collins_2024}, incentivizing learners to unbind their motor constraints, rather than remain locked into a low-dimensional repertoire optimized for known goals. But such calibration presumably requires a means to infer the teleological depth of a given environment. How might this work?

\begin{figure*}[tb]  \centering \includegraphics[width=\textwidth]{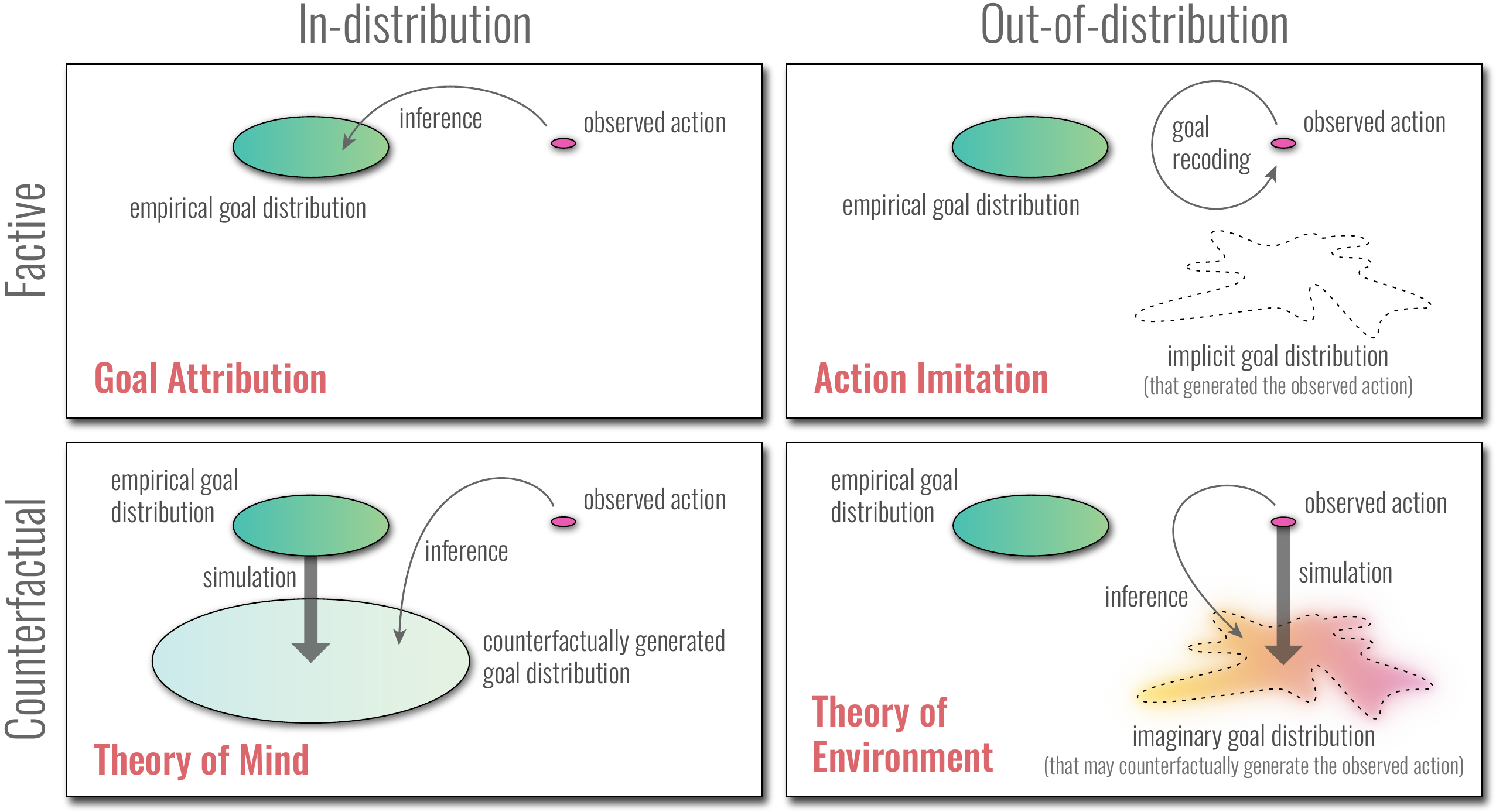} \caption{Four modalities of social goal inference, identifying structural variation along two representational dimensions: (a) \textit{factive} vs. \textit{counterfactual}, pertaining to the omission/use of counterfactual simulation; and (b) \textit{in-} vs. \textit{out-of-distribution}, pertaining to the scope of goal inference -- either bounded or unbounded by a current known hypothesis space.} \label{fig:fig_tax} \end{figure*}

\section{A four-fold typology of social goal inference}  

Echoing \citeauthor{Vygotsky_1980} (1980), we argue that teleological depth is cued by the socio-cultural environment. Without such cued information, the density of latent goals in an environment could only be probed through actual open-ended behavioral exploration -- a prohibitively risky investment, as discussed. We refer to this social inference of teleological depth as \textit{theory of environment} (ToE), and situate it in a 2$\times$2 typology with other better studied mechanisms of social goal inference (\textit{Figure 1}):

\textbf{1. Goal attribution:} From the first year of life, human infants expect others' actions to be goal-directed. Infants are prolific in their attribution of goals not only to observed behaviors, but also to artifactual and natural objects, for example when interpreting the agentic purpose of wrenches or clouds \cite{Kelemen_1999}.

\textbf{2. Theory of Mind (ToM):} When observing an agent who acts upon a false belief, simple goal attribution is thwarted, instead requiring ``meta-representation'' of hidden mental states and counterfactual goals, i.e., theory of mind (ToM). Full-fledged ToM appears later in development than goal attribution \cite{Gergely_Csibra_2003}, and is observed reliably only in humans. Some non-human primates use a simpler, ``factive'' ToM that circumvents the computational cost of counterfactual simulation \cite{Phillips_Buckwalter_Cushman_Friedman_Martin_Turri_Santos_Knobe_2021}. Due in part to this cost of counterfactual use, hypothesis-generation in ToM is constrained to the well-defined (``in-distribution'') space of known goals. This limitation is shared by \textit{inverse reinforcement learning} (IRL) – a common algorithmic approximation of ToM \cite{Baker_Jara-Ettinger_Saxe_Tenenbaum_2017}. IRL scales poorly in complex environments, and is often restricted to closed-ended task domains. ToM is thus inadequate when observing someone posting mail, if the observer lacks prior knowledge of the environmental dynamics of mail service. A cumulatively cultural species is guaranteed regular encounters with such \textit{causally opaque} behavior \cite{Henrich_2016}, suggesting the need for ``out-of-distribution'' inference mechanisms.

\textbf{3. Action Imitation:} Imitation can be seen as the \textit{recoding} of an observed action into a novel goal unto itself \cite{Lyons_Young_Keil_2007, Schachner_Carey_2013} -- mechanistically consistent with \textit{hindsight relabeling} methods in goal-conditioned RL \cite{Andrychowicz_Crow_2017}. Imitation circumvents the cost of counterfactual generation, making it a sample-efficient mechanism for out-of-distribution learning, akin to \textit{episodic control} \cite{Lengyel_Dayan_2008}. But being tethered to literal observations, imitation lacks the generative flexibility of counterfactual simulation.

\textbf{4. Theory of Environment (ToE):} In our mail-posting example, ToM fails to resolve the observed action. But such ``convergence failure'' can itself serve as a valuable cue to switch from postulating hidden mental states to postulating hidden environmental dynamics. Both ToM and ToE depend upon counterfactual generation, and likely draw upon a common computational machinery. But whereas ToM resolves ambiguity by searching a known (in-distribution) hypothesis space, ToE does so by searching an open-ended space of possible environmental dynamics. Verification of possible environmental dynamics requires actual behavioral exploration, rather than internal hypothesis-fitting. ToE thus generates \textit{out-of-distribution counterfactuals} -- a representation adjacent to imagination, which we suggest promotes the expansion of motor dimensionality for skill development. 

In sum: the computational mechanisms of ToM may have purpose beyond mentalization. ToE deploys counterfactual generation of possible worlds, to support open-ended behavioral exploration. Cultural evolutionary theories typically construe asocial and social learning as exploration and exploitation, respectively. Our proposal decomposes this dichotomy by postulating an essential socio-cultural basis for human behavioral exploration and innovation.
\bigskip

\section*{Acknowledgments}
This work is supported by the SUTD Kickstarter Initiative (MOE AcRF Tier 1) under grant number SKI 2021\_06\_09

\bibliography{aaai2026}

\end{document}